%% file: Comm-Eff_SS_with_LK.tex
\documentclass[conference,a4,onecolumn]{IEEEtran}
\usepackage[T1]{fontenc}
\usepackage{fourier}

\usepackage[english]{babel}															
\usepackage[protrusion=true,expansion=true]{microtype}	
\usepackage{amsmath,amsfonts,amsthm} 
\usepackage[pdftex]{graphicx}	
\usepackage{url}

\input{rawad_macros.tex}

\usepackage{fancyhdr}
\IEEEoverridecommandlockouts

\title{
		 Communication Efficient Secret Sharing in the Presence of Malicious Adversary 
}

\author{
\IEEEauthorblockN{
        Rawad~Bitar\IEEEauthorrefmark{2},
        and~Sidharth~Jaggi\IEEEauthorrefmark{3}
        }

\IEEEauthorblockA{ \small
\IEEEauthorrefmark{2} Department of Electrical and Computer Engineering, Rutgers University, Piscataway, NJ 08854, USA \\
\IEEEauthorrefmark{3} Department of Information Engineering, Chinese University of Hong Kong, Shatin, HK, China  \\
Emails: {\tt rawad.bitar@rutgers.edu}, and {\tt jaggi@ie.cuhk.edu.hk}
}
}
\begin{document}
\maketitle

\begin{abstract}
Consider the communication efficient secret sharing problem. A dealer wants to share a secret with $n$ parties such that any $k\leq n$ parties can reconstruct the secret and any $z<k$ parties eavesdropping on their shares obtain no information about the secret. In addition, a legitimate user contacting any $d$, $k\leq d \leq n$, parties to decode the secret can do so by reading and downloading the minimum amount of information needed.


We are interested in communication efficient secret sharing schemes that tolerate the presence of malicious parties actively corrupting their shares and the data delivered to the users. The knowledge of the malicious parties about the secret is restricted to the shares they obtain. We characterize the capacity, \ie maximum size of the secret that can be shared. We derive the minimum amount of information needed to to be read and communicated to a legitimate user to decode the secret from $d$ parties, $k\leq d \leq n$. Error-correcting codes do not achieve capacity in this setting. We construct codes that achieve capacity and achieve minimum read and communication costs for all possible values of $d$. Our codes are based on Staircase codes, previously introduced for communication efficient secret sharing, and on the use of a pairwise hashing scheme used in distributed data storage and network coding settings to detect errors inserted by a limited knowledge adversary.
\end{abstract}

\section{Introduction}
Secret sharing \cite{S79,McESa81} consists of a dealer who wants to share a secret with $n$ parties such that any subset of $z<n$ parties eavesdropping on their shares obtain no information about the secret. Besides its application to privately storing information in a distributed storage system, e.g., \cite{PRR11}, secret sharing is the main tool used in several applications of distributed systems such as private coded computing, e.g., \cite{AF10,BPR17}, private information retrieval \cite{chor1998private} and secure multi-party computations \cite{SMPCbook}. The main challenge arising in distributed systems is tolerating the presence of slow or unresponsive nodes, referred to as stragglers in the distributed computing community \cite{DB13}. Classical secret sharing \cite{S79,McESa81} mitigates the stragglers by allowing a legitimate user to decode the stored information from any subset of nodes of a predetermined size $k$, $z<k\leq n$.

However, in many applications the number of stragglers is not know a priori. Therefore, communication efficient secret sharing is a better fit for this problem. Communication efficient secret sharing (CE-SS) introduced in \cite{WW08} allows a user to decode the secret from any $d$, $k\leq d \leq n$ parties while communicating the minimum amount of information needed to decode the secret. The amount of information communicated to the user using CE-SS is always less than or equal to the amount of information communicated when using classical secret sharing. As a direct application, CE-SS reduces the aggregate delays experienced by the dealer in private coded computing \cite{BPR17}.

We are interested in communication efficient secret sharing schemes that tolerate the presence of malicious parties trying to actively corrupt the data stored in the distributed system. One direct solution is to use CE-SS Reed-Solomon type codes such as the codes introduced in \cite{BRIT18,HLKBtrans}. However, such error-correction codes assume that the malicious parties are omniscient and know the secret stored in the system. This assumption does not hold in this setting where the goal is to maintain the privacy of the dealer's data. This type of adversary is known as limited-knowledge adversary in the literature \cite{PRR11,JLKHKME08}. 

\noindent{\em Related works:} Communication efficient secret sharing problem is introduced in \cite{WW08}. The minimum communication cost as function of the number of stragglers is derived in \cite{WW08,HLKBtrans} and codes achieving the bound are given in \cite{WW08,ZYSMH12,BRIT18,HLKBtrans}. On the other hand, protecting distributed systems from limited-knowledge malicious adversary is studied in different settings. In \cite{JLKHKME08,yao2014network} the authors consider a network coding setting in which the adversary can corrupt the data sent through some of the network's nodes and derive the capacity of such system. Reliable distributed storage system under repair dynamics is studied in \cite{PRR11,bitar2015securing}. The authors of \cite{PRR11} derive the capacity of reliable and secure distributed storage systems under the repair dynamics. Capacity-achieving codes are provided in \cite{PRR11,bitar2015securing}. The introduced codes are based on codes for non-reliable storage systems coupled with the pairwise hashing scheme introduced in \cite{yao2014network}. In both settings, it is shown that leveraging the limitation of the adversary's knowledge leads to increasing the capacity of the system \cite{PRR11,JLKHKME08,bitar2015securing,schaefer2017information}.

\noindent{\em Contributions:} We study the problem of reliable communication efficient secret sharing. We derive the capacity of such systems, \ie the maximum size of the secret that can be reliably and privately stored in a distributed system in the presence of an unknown number of stragglers and a limited-knowledge adversary eavesdropping on the data and actively trying to corrupt the stored data. We characterize the minimum amount of information that the user has to download to decode the secret as a function of the number of stragglers, number of eavesdropped nodes, and number of corrupted nodes. We provide codes that achieve capacity and the minimum communication cost for any number of stragglers. Our codes are based on the use of Staircase codes and a pairwise hashing scheme that allows the user to detect the malicious parties. We compare the derived capacity to the error-correction capacity of the system with privacy constraints, \ie assuming the adversary corrupting the contents of the shares is omniscient. As a result, we show that leveraging the limitation of the adversary's knowledge increases the capacity of the system. We illustrate the ideas in the following Example.

\begin{example}
We construct a reliable communication efficient secret sharing with $n=4$ parties. Assume that at most $1$ party can be a straggler, \ie $k=3$ and that the adversary can spy and corrupt the content of $1$ party, \ie $z=1$. Let $\bs=(\bs_1,\bs_2) \in  \F_q^2$, $q\geq 4,$ be the secret to be stored. The dealer generates two random numbers $\br_1$ and $\br_2$ drawn independently and uniformly at random from $\F_q$ and independently from $\bs$. The shares given to the parties are computed using Staircase codes \cite{BRIT18} and are shown in table~\ref{tab:cess}.

For each party $i$ we denote by $\bw_{i1}$ and $\bw_{i2}$ the first and second part of the share given to that party. We view $\bw_{i1}$ and $\bw_{i2}$ as vectors in some finite field  $\F_{q_1}^v$ and denote by $\langle\bw_{i1},\bw_{j1}\rangle$ the dot product of two vectors $\bw_{i1}$ and $\bw_{j1}$. The dealer computes the following pairwise hashes $\bh_{i1} = (\langle\bw_{i1},\bw_{j1}\rangle)$ for all $i \neq j \in \{1,\dots,4\}$ and $\bh_{i2} = (\langle\bw_{i2},\bw_{j2}\rangle)$ for all $i\neq j\in \{1,\dots,4\}$ and sends $\bh_{i1}$ and $\bh_{i2}$ to party $i$.

\begin{table}[t!]
\centering
\begin{tabular}[h!]{c|c|c|c}
Party 1& Party 2 & Party 3 & Party 4\\ \hline
\blue $\bs_1+\bs_2+ \br_1$ & \blue $\bs_1+2\bs_2+4\br_1$& \blue $\bs_1+3\bs_2+ 4\br_1$& \blue $\bs_1+4\bs_2+\br_1$ \\
$\br_1+\br_2$ & $\br_1+2\br_2$ & $ \br_1+3\br_2$ & $ \br_1+4\br_2$ \\
\end{tabular}
\caption{The Staircase secret sharing code for $n=4$, $k=3$, $z=1$ and achieves minimum communication cost for $d'=2$ and $d'=3$ over $\F_5^2$.}
\label{tab:cess}
\end{table}

Without loss of generality assume that party~$1$ is controlled by the adversary. A user contacting $d=4$ parties downloads the first half of each share and all the hashes $\bh_{i1}$. Assuming that the size of the hash is negligible compared to the size of the shares, the communication cost is equal to $4$ units of information. Note that the adversary only observes the share of party~$1$ and therefore the other shares are uniformly distributed over $\F_{q_1}^{v}$ from his perspective. The user computes $\hat{\bh}_{i1}$ for $i=1,2,3,4$ and compares them to the downloaded hash $\bh_{i1}$. The only corrupted packet here is $\hat{\bw}_{11}$ and can be written as $\hat{\bw}_{11}={\bw_{11}}+\mathbf{e}$ for a given error vector $\mathbf{e}$. Since $\bw_{i1}$ is independent from the adversary's observation, therefore
\begin{align*}
\Pr(\hat{\bh}_{11}  = \bh_{11}) &= \Pr(\mathbf{e} \bot \bw_{21}, \mathbf{e} \bot \bw_{31}, \mathbf{e} \bot \bw_{41})\\
& = \Pr(\mathbf{e} \bot \bw_{21}) \Pr( \mathbf{e} \bot \bw_{31}|\mathbf{e} \bot \bw_{21}) \Pr( \mathbf{e} \bot \bw_{41}|\mathbf{e} \bot \bw_{21},\mathbf{e} \bot \bw_{31})\\
& \leq \Pr(\mathbf{e} \bot \bw_{21})\\
& = \dfrac{1}{q^{v}}.
\end{align*}

With high probability, the user can construct the following hash comparison table where $\times$ denotes that $\hat{\bh}_{i1}\neq \bh_{i1}$ and $\checkmark$ denotes equality. The user looks at the row with the most number of $\times$ and declares the party corresponding to this row as corrupted. Note that if the first row had just one $\times$, say $\hat{\bh_{11}}\neq \bh_{21}$, the user cannot know whether party $1$ or party $2$ is corrupted. This happens with probability at most $q^{-v}$.

\begin{table}[h!]
\centering
\begin{tabular}{c|c|c|c|c}
~ & $\hat{\bh}_{11}$ & $\hat{\bh}_{21}$ & $\hat{\bh}_{31}$ & $\hat{\bh}_{41}$\\ \cline{1-5}
${\bh_{11}}$ & $\checkmark$ & $\times$ & $\times$ & $\times$ \\ \hline
${\bh_{21}}$ & $\times$ & $\checkmark$ & $\checkmark$ & $\checkmark$\\ \hline
${\bh_{31}}$ & $\times$ & $\checkmark$ & $\checkmark$ & $\checkmark$\\ \hline
${\bh_{41}}$ & $\times$ & $\checkmark$ & $\checkmark$ & $\checkmark$\\
\end{tabular}
\end{table}

The user deletes the data downloaded from party~$1$ and decodes the secret using the other downloaded shares. Similarly, a user contacting any $d=k=3$ parties downloads all their shares and does the same as above. The communication cost is $6$ units of information and the probability of error is upper bounded by $q^{-v}$.

In this example, the size of the secret is equal to $2$ symbols. A user contacting $d=4$ parties reads and downloads $4$ units of information and a user contacting $d=3$ parties reads and downloads $6$ units of information. The user can detect the corrupted node and decode the secret with high probability. We show in the sequel that this code is optimal, i.e., the secret size achieves capacity and the minimum download costs for $d=4$ and $d=3$ are $4$ and $6$ units of information, respectively.

Note that the error correction capacity of the system, \ie if the adversary were omniscient, reduces to $0$. From the singleton bound we know that the amount of information that can be reliably stored in the system is equal to one share. However, this is exactly the amount of randomness needed to maintain privacy of the data, and therefore the secret can be of size $0$.
\end{example}

\section{Problem Formulation}
We consider the problem of communication efficient secret sharing in the presence of a malicious adversary. In classical secret sharing setting, the dealer wants to share a secret $\bs$ with $n$ parties such that a user can decode the secret from any subset of $k\leq n$ parties by downloading all their shares. In addition, any subset of $z_r<k$ parties should not obtain any information about the secret. We assume that the share given to each party consists of $\alpha$ symbols each being an element of a finite field $\F_q$, where $q\geq n$ is a power of a prime. For the scheme to be communication efficient, we require that a user contacting $d$, $k\leq d\leq n$ parties can decode the secret by downloading less then $k$ shares. The minimum communication cost as function of $d$ is given by \cite{WW08,HLKBtrans} $\CC(d) = d\dfrac{(k-z_r)\alpha}{d-z_r}$. This implies that the user can tolerate the presence of $n-d$ stragglers for $k\leq d \leq n$.

The new constraint that we impose here is that up to $z_w$ parties can be malicious and can send corrupted data to a user reconstructing the secret. The adversary James has different control level on the parties. It can eavesdrop on $z_{ro}$ parties, blindly corrupt (jam) the content of $z_{wo}$ parties, and eavesdrop and corrupt the content of $z_{rw}$ parties. Let $\mathbf{z} \triangleq (z_{ro}, z_{wo},z_{rw})$. Note that by definition $z_r=z_{ro}+z_{rw}$ and $z_w = z_{wo} + z_{rw}$. We study distributed storage systems that satisfy the following properties. 

\paragraph{Perfect privacy}
Let $\mathbf{s}$ be the secret and let $S$ be the random variable denoting the secret. Let $\mathbf{w}_i$ be the share given to party $i$ and let $W_i$ denote the random variable representing $\mathbf{w}_i$. For any set $\setB \subset \{1,\dots,n\}$, let $W_\setB$ denote the shares given to the parties indexed by $\setB$, \ie $W_\setB =\{W_i; i\in \setB\}$. The privacy constraint is expressed as
\begin{equation}\label{eq:privacy}
H(S|W_\setZ) = H(S), \quad \forall \setZ \subset [n], |\setZ|=z_r.
\end{equation}
Here $H$ is the entropy function and all logarithms are base $q$.
\paragraph{Resiliency}
A user contacting $k$ parties and downloading all their shares can decode the secret. The resiliency requirement can be expressed as 
\begin{equation}\label{eq:resiliency}
H(S|W_\setA) = 0, \quad \forall \setA \subseteq [n], |\setA|=k.
\end{equation}

Let $\hat{\mathbf{s}}$ be the secret reconstructed by the user. We relax the condition of zero-error reconstruction and allow a small probability of error, \ie for all $\varepsilon>0$, $\Pr_e \triangleq \Pr \left( \hat{\mathbf{s}} \neq \bs \right) < \varepsilon$.

\paragraph{Communication efficiency}
A user contacting any $d$ parties, $k\leq d \leq n$, decodes the secret by reading and downloading the minimum amount of information. 

Note that the read cost is upper bounded by the communication cost since the parties must at least read the amount of information communicated to the user.

In the sequel, we denote by $(n,k,\mathbf{z})$ a reliable communication efficient secret sharing (R-CE-SS) with $n$ parties, a threshold on the stragglers equal to $n-k$ and an adversary controlling $\bz$ parties as defined above. Our goal is to derive the capacity of an $(n,k,\mathbf{z})$ R-CE-SS. In other words, we want to find the maximum size of the secret $\bs$ that can be stored in the distributed system using an $(n,k,\mathbf{z})$ R-CE-SS and provide codes that achieve capacity and minimum download cost for all number of stragglers. We consider two types of adversaries: limited knowledge adversaries and omniscient adversaries. The former adversary only observes the information shared with $z_r$ parties and corrupt the content of $z_w$ shares. Whereas, the latter has full knowledge of all the shares information and can corrupt the content of $z_w$ shares. We consider omniscient adversary to model the worst case error correction capacity.

\section{Main Results}
\noindent{\em Limited knowledge adversary:} We characterize the capacity $C^\text{LK}(n,k,\bz)$ of an $(n,k,\bz)$ R-CE-SS scheme and derive the minimum communication and read costs, $\CC(d)$ and $\RC(d)$ incurred by a user contacting $d$ parties to decode the secret.

\begin{theorem}\label{thm:main}
The capacity of an $(n,k,\bz)$ R-CE-SS, where $\bz= (z_{ro}, z_{wo},z_{rw})$,  in the presence of a limited knowledge adversary eavesdropping on $z_{ro}$ shares, blindly corrupting the content of $z_{wo}$ shares and eavesdropping and corrupting the content of $z_{rw}$ shares is given by
\begin{equation}\label{eq:capacity}
C^\text{LK}(n,k,\bz) = \begin{cases}
\hfill (k-z_r-z_w)\alpha & \hfill \text{if    }  k> 2z_{rw}+2z_{wo} + z_{ro},\\
\hfill 0 & \hfill \text{otherwise}.
\end{cases}
\end{equation}

The communication and read costs $\CC(d)$ and $\RC(d)$ incurred by a user contacting $d$ parties to decode the secret are
\begin{equation}\label{eq:cc}
\RC(d) = \CC(d) = d\dfrac{(k-2z_{rw}-z_{wo}-z_{ro})\alpha}{d-2z_{rw}-z_{wo}-z_{ro}}.
\end{equation}
\end{theorem}

We construct R-CE-SS codes that achieve $C^\text{LK}(n,k,\bz)$ and the minimum costs simultaneously for all values of $d$, $k\leq q\leq n$. Our codes are based on the use of Staircase codes \cite{BRIT18} and pairwise hash \cite{yao2014network,PRR11}.

\vspace{0.2cm}

\noindent{\em Omniscient adversary:} We derive the capacity $C^\text{O}(n,k,\bz)$ of an $(n,k,\bz)$ R-CE-SS scheme to show that  $C^\text{LK}(n,k,\bz) > C^\text{O}(n,k,\bz)$. The capacity in this setting follows from Singleton-type bounds with privacy constrains. We also derive the minimum communication and read costs, $\CC(d)$ and $\RC(d)$ incurred by a user contacting $d$ parties to decode the secret.

\begin{theorem}\label{thm:main}
The capacity of an $(n,k,\bz)$ R-CE-SS, where $\bz= (z_{ro}, z_{wo},z_{rw})$, in the presence of an adversary eavesdropping on $z_{ro}+z_{rw}$ shares and an omniscient adversary corrupting the content of $z_{wo}+z_{rw}$ shares is given by
\begin{equation}\label{eq:capacity}
C^\text{O}(n,k,\bz) = \begin{cases}
\hfill (k-z_r-2z_w)\alpha & \hfill \text{if    }  k> 3z_{rw}+2z_{wo} + z_{ro},\\
\hfill 0 & \hfill \text{otherwise}.
\end{cases}
\end{equation}

The communication and read costs $\CC(d)$ and $\RC(d)$ incurred by a user contacting $d$ parties to decode the secret are
\begin{equation}\label{eq:cc}
\RC(d) = \CC(d) = d\dfrac{(k-3z_{rw}-2z_{wo}-z_{ro})\alpha}{d-3z_{rw}-2z_{wo}-z_{ro}}.
\end{equation}
\end{theorem}

Staircase codes \cite{BRIT18} and the codes presented in~\cite{HLKBtrans} are R-CE-SS codes that achieve $C^\text{O}(n,k,\bz)$ and the minimum costs simultaneously for all values of $d$, $k\leq q\leq n$.

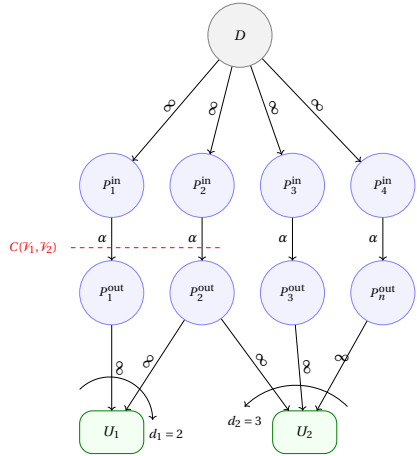
\begin{figure}[b!]
\centering
\resizebox{!}{6cm}{
\begin{tikzpicture}
    \tikzstyle{stealth} = [draw=none,text=black]
    \tikzstyle{source}  = [draw=gray,fill=black!5,circle, minimum width=1.5cm]
    \tikzstyle{storage} = [draw=blue!50,fill=blue!5,thin, circle, minimum width=1.5 cm,font=\small]
    \tikzstyle{user} = [draw=green!50!black, thin, minimum width =1.5cm, minimum height = 1cm, rounded corners = 3mm, fill=green!5]
    \tikzstyle{snode} = [draw=blue!50, thin, minimum width =1.5cm, minimum height= 0.8cm, fill=blue!5,rounded corners=3mm]
    \tikzstyle{rnode} = [draw=red!50, dashed,thin, minimum width =1.5cm, minimum height= 0.8cm, fill=red!5,rounded corners=3mm]

\node[source] at (0,0) (d) {$D$};

\node[storage, below = 2cm of d, xshift = -3cm] (p1i) {$P_1^\text{in}$};
\node[storage, right = 0.6cm of p1i] (p2i) {$P_2^\text{in}$};
\node[storage, right = 0.6 cm of p2i] (p3i) {$P_3^\text{in}$};
\node[storage, right = 0.6 cm of p3i] (p4i) {$P_4^\text{in}$};

\node[storage, below = 1cm of p1i] (p1o) {$P_1^\text{out}$};
\node[storage, right = 0.6cm of p1o] (p2o) {$P_2^\text{out}$};
\node[storage, right = 0.6 cm of p2o] (p3o) {$P_3^\text{out}$};
\node[storage, right = 0.6 cm of p3o] (p4o) {$P_n^\text{out}$};

\node[user, below = 2cm of p1o] (u1) {$U_1$};
\node[user, right = 3cm of u1] (u2) {$U_2$};

\path[->]
(d) edge node[above, midway, sloped] {$\infty$} (p1i)
edge node[above, midway, sloped] {$\infty$} (p2i)
edge node[above, midway, sloped] {$\infty$} (p3i)
edge node[above, midway, sloped] {$\infty$} (p4i);

\path[->]
(p1i) edge node[left, midway] (a1) {$\alpha$} (p1o)
(p2i) edge node[left, midway] (a2) {$\alpha$} (p2o)
(p3i) edge node[left, midway] {$\alpha$} (p3o)
(p4i) edge node[left, midway] {$\alpha$} (p4o);

\path[->]
(p1o) edge node[above, midway, sloped] {$\infty$} (u1)
(p2o) edge node[above, midway, sloped] {$\infty$} (u1)

(p2o) edge node[above, midway, sloped] {$\infty$} (u2)
(p3o) edge node[above, midway, sloped] (i1) {$\infty$} (u2)
(p4o) edge node[above, midway] (i2) {$\infty$} (u2)
;

  \draw[->,thin] (u1) [yshift = 1cm, xshift = -0.75cm] arc (135:0:1cm) node[below right=0.1cm and -0.2cm] {\small $d_1=2$};
  \draw[->,thin] (u2) [yshift = 0.6cm, xshift = 1cm] arc (45:135:1.7cm) node[below=0.1cm] {\small $d_2=3$};
      
\node[below left = -0.1cm and 0.4 cm of a1] (a1){};
\node[below right = -0.1cm and 0.4 of a2] (a2) {};      
      
    \draw[red, dashed] plot [smooth, tension=0.5] coordinates {(a1) (a2)};
        \node[stealth,red, left = 0.1 of a1] {\small $C(\setV_1,\setV_2)$};

\end{tikzpicture}
}
\caption{A depiction of the graph representation of the reliable communication efficient secret sharing problem with $n=4$ workers and two values of $d$, $d_1=2$ and $d_2 = 3$. The dealer is the source represented by a vertex $D$. Each party $i$ is represented by two vertices $P_i^\text{in}$ and $P_i^\text{out}$. A user $j$ is a terminal $U_j$ connected to $d$ parties for different values of $d$. All edges are directed and the capacity is mentioned on the edge. A cut between the dealer and user $1$ is shown where $\setV_2 =\{P_1^\text{out}, P_2^\text{out}, U_1\}$ and $\setV_1 = \setV \setminus \setV_2$. The value of this cut is $2\alpha$.}
\label{fig:network}
\end{figure}

\section{Flow Graph Representation}
We look at the problem at hand as a multicast problem on parallel edges. We model the network using a flow graph $\mathcal{G}(\mathcal{V},\mathcal{E})$ with $\mathcal{V}$ being the set of vertices and $\mathcal{E}$ the set of edges. This representation is introduced in \cite{DGWWR07}. The dealer ${D}$ is the source of the network and each user $j$ is a terminal ${U_j}$. We view each party as being two vertices in the graph, a vertex $\mathcal{P}^\text{in}$ and $\mathcal{P}^\text{out}$. The vertex $D$ is connected to $P_i^\text{in}$ with an edge $D \to P_i^\text{in}$ of infinite capacity for all $i = 1,\dots,n$. For all $i = 1,\dots,n$, the vertex  $P_i^\text{out}$ is connected to $P_i\text{in}$ with an edge $P_i^\text{in} \to P_i^\text{out}$ of capacity equal to $\alpha$, the storage capacity of each party. A vertex $U_j$ is connected to any set of $d$ vertices $P_i^\text{out}$, indexed by $\mathcal{I}_j\subseteq [n]$, with edges of infinite capacity $P_i^{out} \to U_j$ for all $i\in \setI_j$. We depict this network in Figure~\ref{fig:network}.

A cut $C(\setV_1,\setV_2)$ between $D$ and a given user $U_j$ in the network is defined as a partition of the set of vertices $\setV$ into two sets $\setV_1$ and $\setV_2$ such that:
\begin{enumerate}
\item By definition of a partition, $\setV= \setV_1 \cup \setV_2$ and $\setV_1 \cap \setV_2 = \emptyset$.
\item The source $D$ is in $\setV_1$ and $U_j$ is in $\setV_2$.
\item There is no edge in the network that connects any vertex in $\setV_1$ to the user $U_j$.
\end{enumerate}

The value of a cut $C(\setV_1,\setV_2)$ is defined as the sum of the capacities of the edges going from a vertex in $\setV_1$ to a vertex in $\setV_2$.  The idea is to use the min-cut max-flow to bound the capacity of the network. For example, in the network shown in Figure~\ref{fig:network} the dealer can send at most $2\alpha$ units of information to user $1$. This amount is equal to the value of the cut $C(\setV_1,\setV_2)$.

\section{Converses}
Let $\alpha$ be the amount of information given to each party. We show that $H(S)\leq (k- 2z_{rw} - z_{ro} - z_{wo})\alpha$ using standard information theoretic inequalities. Then, we use the flow graph information representation of the reliable communication efficient secret sharing system to show the following.
\begin{equation}\label{eq:capacity}
C(n,k,\bz) \leq \begin{cases}
\hfill k-2z_{rw}-z_{wo} - z_{ro} & \hfill \text{if    } k> 2z_{rw}+2z_{wo} + z_{ro},\\
\hfill 0 & \hfill \text{otherwise}.
\end{cases}
\end{equation}

\subsection{Bound on the Entropy of the Secret}\label{sec:bound}
Let $W_i^j$ $, i<j$, denote the set of random variables $W_i, \dots, W_j$ and recall that $z_w = z_{wo}+ r_{rw}$ and $z_r = z_{ro}+ r_{rw}$. For any collection of $k$ shares we can write the following using~\eqref{eq:resiliency}
\begin{align}
H(S) &=  I(S; W_1^k) \label{eq:mutual}\\
 & = I(S; W_1^{z_{w}}) + I(S; W_{z_{w}+1}^{k}| W_1^{z_{w}}) \label{eq:chain}\\
 & = I(S; W_{z_{w}+1}^{k}| W_1^{z_{w}}) \label{eq:pri}\\
 & = H(W_{z_{w}+1}^{k}| W_1^{z_{w}}) - H(W_{z_{w}+1}^{k} | S,W_1^{z_{w}}) \nonumber\\
 & = H(W_{z_{w}+z_{r}+1}^{k}| W_1^{z_{w}}) \nonumber \\
 & ~~ +  H(W_{z_{w}+1}^{z_{w}+z_{r}}| W_1^{z_{w}}, W_{z_{w}+z_{r}+1}^{k}) - H(W_{z_{w}+1}^{k} | S,W_1^{z_{w}}) \label{eq:chain2} \\
 & \leq H(W_{z_{w}+z_{r}+1}^{k}) \label{eq:dpi} \\
 & \leq (k-2z_{rw}-z_{wo} - z_{ro})H(W_i) \label{eq:ind}\\
 & = (k-2z_{rw}-z_{wo} - z_{ro})\alpha. \label{eq:alpha}
\end{align}

Equation~\eqref{eq:chain} follows from the chain rule of mutual information. Equation~\eqref{eq:pri} follows from the privacy constraint given in~\eqref{eq:privacy}. In~\eqref{eq:pri} we removed the first $z_{ro}$ shares which does not incur loss of generality. Equation~\eqref{eq:chain2} follows from the chain rule of entropy. Equation~\eqref{eq:dpi} follows from the data processing inequality as we shall show next. Equation~\eqref{eq:ind} follows from the chain rule of entropy and~\eqref{eq:alpha} follows because $H(W_i) = \alpha$. 

To show that~\eqref{eq:dpi} holds we use the non-negativity of the entropy and write the following.
\begin{align}
H(W_{z_{w}+1}^{k} | S,W_1^{z_{w}}) & = H(W_{z_{w}+1}^{z_{w}+z_{r}}| S,W_1^{z_{w}}) +   H(W_{z_{w}+z_{r}}^{k}| S,W_1^{z_{w}}, W_{z_{w}+1}^{z_{w}+z_{r}}) \nonumber \\
& \geq H(W_{z_{w}+1}^{z_{w}+ z_{r}}| S,W_1^{z_{w}}).\label{eq:greater}
\end{align}

Note that~\eqref{eq:greater} holds with equality because all the shares are a deterministic function of the secret and the randomness that can be extracted from any $z_r$ shares. Let $$Q\triangleq H(W_{z_{w}+1}^{z_{w}+z_{r}}| W_1^{z_{w}},W_{z_{w}+z_{r}+1}^{k}) - H(W_{z_{w}+1}^{k} | S,W_1^{z_{w}}),$$ we use~\eqref{eq:greater} to bound $Q$ as follows.
\begin{align}
Q &\leq H(W_{z_{w}+1}^{z_{w}+z_{r}}| W_1^{z_{w}},W_{z_{w}+ z_{r}+1}^{k}) - H(W_{z_{w}+ 1}^{z_{w}+z_{r}}| S,W_1^{z_{w}}) \nonumber \\
& =I(W_{z_{w}+1}^{z_{w}+z_{r}}; S,W_1^{z_{w}}) - I(W_{z_{w}+1}^{z_{w}+z_{r}}; W_1^{z_{w}},W_{z_{w}+z_{r}+1}^{k})\label{eq:defi}\\
& \geq 0.\label{eq:dpi1}
\end{align}

Equation~\eqref{eq:defi} follows from the definition of mutual information $I(A;B) = H(A) - H(A|B)$ and~\eqref{eq:dpi1} holds because $S\to W_{z_{w}+z_{r}+1}^{k}$ forms a Markov chain and we can use the data processing inequality.

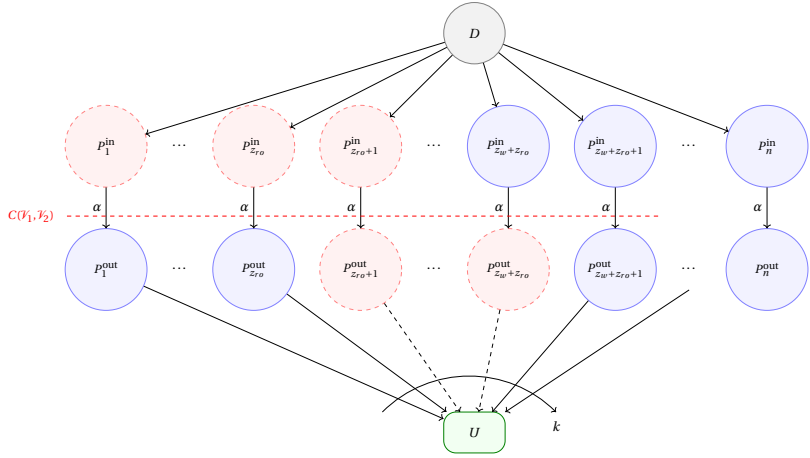
\begin{figure}[t!]
\centering
\resizebox{!}{6cm}{
\begin{tikzpicture}
    \tikzstyle{stealth} = [draw=none,text=black]
    \tikzstyle{source}  = [draw=gray,fill=black!5,circle, minimum width=1.5cm]
    \tikzstyle{storage} = [draw=blue!50,fill=blue!5,thin, circle, minimum width=2 cm,font=\small]
    \tikzstyle{user} = [draw=green!50!black, thin, minimum width =1.5cm, minimum height = 1cm, rounded corners = 3mm, fill=green!5]
    \tikzstyle{rwstorage} = [draw=red!50,fill=red!5,thin, circle, dashed, minimum width=2 cm,font=\small]

\node[source] at (0,0) (d) {$D$};

\node[rwstorage, below = 1cm of d, xshift = -9cm] (p1i) {$P_1^\text{in}$};
\node[right = 0.5 cm of p1i] (dots) {$\cdots$};
\node[rwstorage, right = 0.5cm of dots] (p2i) {$P_{z_{ro}}^\text{in}$};
\node[rwstorage, right = 0.6 cm of p2i] (p3i) {$P_{z_{ro}+1}^\text{in}$};
\node[right = 0.5 cm of p3i] (dots) {$\cdots$};
\node[storage, right = 0.5 cm of dots] (p4i) {$P_{z_{w}+z_{ro}}^\text{in}$};
\node[storage, right = 0.6 cm of p4i] (p5i) {$P_{z_{w}+z_{ro}+1}^\text{in}$};
\node[right = 0.5 cm of p5i] (dots) {$\cdots$};
\node[storage, right = 0.6 cm of dots] (pni) {$P_n^\text{in}$};

\node[storage, below = 1cm of p1i] (p1o) {$P_1^\text{out}$};
\node[right = 0.5 cm of p1o] (dots) {$\cdots$};
\node[storage, right = 0.5cm of dots] (p2o) {$P_{z_{ro}}^\text{out}$};
\node[rwstorage, right = 0.6 cm of p2o] (p3o) {$P_{z_{ro}+1}^\text{out}$};
\node[right = 0.5 cm of p3o] (dots) {$\cdots$};
\node[rwstorage, right = 0.5 cm of dots] (p4o) {$P_{z_{w}+z_{ro}}^\text{out}$};
\node[storage, right = 0.6 cm of p4o] (p5o) {$P_{z_{w}+z_{ro}+1}^\text{out}$};
\node[right = 0.5 cm of p5o] (dots) {$\cdots$};
\node[storage, right = 0.6 cm of dots] (pno) {$P_n^\text{out}$};

\node[user, below = 8.5cm of d] (u1) {$U$};

\path[->]
(d) edge node[above, midway, sloped] {} (p1i)
edge node[above, midway, sloped] {} (p2i)
edge node[above, midway, sloped] {} (p3i)
edge node[above, midway, sloped] {} (p4i)
edge node[above, midway, sloped] {} (p5i)
edge node[above, midway, sloped] {} (pni);

\path[->]
(p1i) edge node[left, midway] (a1) {$\alpha$} (p1o)
(p2i) edge node[left, midway] {$\alpha$} (p2o)
(p3i) edge node[left, midway] {$\alpha$} (p3o)
(p4i) edge node[left, midway] {$\alpha$} (p4o)
(p5i) edge node[left, midway] (a2) {$\alpha$} (p5o)
(pni) edge node[left, midway] {$\alpha$} (pno);

\path[->]
(p1o) edge node[above, midway, sloped] {} (u1)
(p2o) edge node[above, midway, sloped] {} (u1)
(p3o) edge [dashed] node[above, midway, sloped] {} (u1)
(p4o) edge [dashed] node[above, midway, sloped] {} (u1)
(p5o) edge node[above, midway, sloped] {} (u1)
(dots) + (0,-0.5) edge node[above, midway, sloped] {} (u1)

;

  \draw[->,thin] (u1) [yshift = 0.5cm, xshift = -2.25cm] arc (135:45:3cm) node[below right=0.1cm and -0.2cm] {$k$};
      
\node[below left = -0.1cm and 0.4 cm of a1] (a1){};
\node[below right = -0.1cm and 1 of a2] (a2) {};      
      
    \draw[red, dashed] plot [smooth, tension=0.5] coordinates {(a1) (a2)};
        \node[stealth,red, left = 0.1 of a1] {\small $C(\setV_1,\setV_2)$};

\end{tikzpicture}
}
\caption{Flow graph representation of the reliable communication efficient secret sharing problem. Edges with no capacity are infinite capacity edges. A dashed $P_i^\text{in}$ means that the adversary can read the share of party $i$.A dashed $P_i^\text{out}$ means that the adversary can change the data sent from party $i$ to the user. We split the parties into $3$ sets, a read-only set (on the left), a write set (in the middle) and a set where the adversary has no control (on the right). The value of the considered cut is $k\alpha$.}
\label{fig:network2}
\end{figure}

\subsection{Limited Knowledge Adversary}
\paragraph{Capacity} We quantify the amount of information that the dealer can send to a user contacting any $k$ parties. We do so by finding a cut in the network between the dealer and such a user. By the min-cut max-flow argument the capacity of the system is upper bounded by the value of this cut. We partition the contacted parties into three disjoint sets. Let $\setR$, $\setW$ and $\setH$ be three disjoint subsets of $[n]$ such that $|\setR|=z_{ro}$, $|\setW|=z_{rw}+z_{wo}$ and $|\setH|=k-z_{ro} - z_{wo} - z_{rw}$. We denote by $P_{\setR}$, $P_{\setW}$, $P_{\setH}$ the set of parties indexed by $\setR$, ${\setW}$ and ${\setH}$, respectively. Let $P_{\setR}$ be the set of parties on which the adversary can eavesdrop (read only), $P_{\setW}$ be the set of parties which shares can be corrupted by the adversary (write only and read write) and  $P_{\setH}$ be the set of parties not controlled by the adversary. A cut between the dealer and the user is $\setV_2 = \{U,P_{\setR}^\text{out}, P_{\setW}^\text{out}, P_{\setH}^\text{out}\}$ and $\setV_2 = \setV \setminus \setV_1$. The value of this cut is $k\alpha$. We divide the set of outgoing edges to the user into three sets (see Figure~\ref{fig:network2} for a pictorial representation):
\begin{enumerate}
\item $\setE_1$: the set of outgoing edges from the parties indexed by $\setR$, \ie the set of edges $P_i^\text{in}\to P_i^\text{out}$ for all $i\in \setR$.
\item $\setE_2$: the set of outgoing edges from the parties indexed by $\setW$, \ie the set of edges $P_i^\text{in}\to P_i^\text{out}$ for all $i\in \setW$.
\item $\setE_3$: the set of outgoing edges from the parties indexed by $\setH$, \ie the set of edges $P_i^\text{in}\to P_i^\text{out}$ for all $i\in \setH$.
\end{enumerate}

Let $X_{\setE_i}(\bs)$ be the symbols sent to the user on the set of edges $\setE_i$ when a secret $\mathbf{s}$ is being transmitted from the dealer. The information sent to the user on $P_i^\text{out}\to U$ depends only on the information sent on $P_i^\text{in}\to P_i^\text{out}$. Therefore we focus on the links $P_i^\text{in}\to P_i^\text{out}$. We consider two cases depending on the value of $k$: %
\begin{enumerate*}[label={\textit{(\roman*)}}]
\item $k \leq 2z_{rw} + 2 z_{wo} + z_{ro}$; and
\item $k > 2z_{rw} + 2 z_{wo} + z_{ro}$.
\end{enumerate*}

\noindent{\em Case 1:} First consider $k = 2z_{rw} + 2 z_{wo} + z_{ro}$. The adversary James can decide on a given secret $\bs_{J}$ of his choice independently from the true secret being sent by the dealer and send $X_{\setE_2}(\bs_J)$ on the set of links he controls. The user now observes $X_{\setE_1}(\bs)$, $X_{\setE_2}(\bs_J)$ and $X_{\setE_3}(\bs)$. From the user's perspective all secrets are equally likely because the dealer is choosing a secret uniformly at random and sending it to the user. Using the upper bound on $H(S)$ we can verify that the uncertainty of the message to the user and to the adversary is the same and is equal $z_{wo}\alpha$. Therefore, since $|\setE_2| = |\setE_3|$ and both $X_{\setE_2}(\bs_J)$ and $X_{\setE_3}(\bs)$ are consistent with $X_{\setE_1}(\bs)$ and both uniformly distributed over the same alphabet the user cannot decide wether $\bs$ or $\bs_J$ is the true message and must decode the message using only $X_{\setE_1}(\bs)$. We formalize this intuition as follows. We assume that the secret $\bs$ is uniformly distributed over $\F_q^{z_{wo}\alpha}$. Let $\br$ and $\br_J$ be the values of the random numbers chosen by the dealer and James respectively. We can write the following.
\begin{align}
\Pr \left( X_{\setE_2}(\bs_J) = x_2, X_{\setE_3}(\bs) = x_3  | X_{\setE_1}(\bs) = x_1 \right) & = \Pr \left (X_{\setE_3}(\bs) = x_3\right) \Pr \left (X_{\setE_2}(\bs_J) = x_2| X_{\setE_1}(\bs) = x_1, X_{\setE_3}(\bs) = x_3\right) \nonumber \\
& = \Pr \left (X_{\setE_3}(\bs) = x_3\right) \Pr \left (X_{\setE_2}(\bs_J) = x_2| X_{\setE_1}(\bs) = x_1\right) \label{eq:nob}\\
& = \Pr \left(S = \bs, R=\br   \right) \Pr \left(S = \bs_J, R=\br_J | X_{\setE_1}(\bs) = x_1   \right) \nonumber \\
& = \Pr \left(S = \bs \right)  \Pr \left(R = \br \right) \Pr \left(S = \bs_J\right) \Pr \left(R = \br_J| X_{\setE_1}(\bs) = x_1   \right) \label{eq:secret_random} \\
& = \dfrac{1}{q^{z_{wo}\alpha}}  \Pr \left(R = \br \right) \dfrac{1}{q^{z_{wo}\alpha}}  \Pr \left(R = \br_J| X_{\setE_1}(\bs) = x_1   \right) \label{eq:secret_uni}\\
& =  \Pr \left(S = \bs_J \right)  \Pr \left(R = \br \right) \Pr \left(S = \bs \right) \Pr \left(R = \br_J| X_{\setE_1}(\bs) = x_1   \right) \nonumber \\
& = \Pr \left (X_{\setE_3}(\bs_J) = x_3\right) \Pr \left (X_{\setE_2}(\bs) = x_2| X_{\setE_1}(\bs) = x_1\right) \nonumber \\
& = \Pr \left( X_{\setE_2}(\bs) = x_2, X_{\setE_3}(\bs_J) = x_3  | X_{\setE_1}(\bs) = x_1 \right). \nonumber
\end{align}

Equation~\eqref{eq:nob} holds because the adversary does not observe the information sent on $\setE_3$. Equation~\eqref{eq:secret_random} follows because the dealer chooses the random numbers independently from the secret to ensure privacy. In addition, the information on sent on $\setE_1$ contains no information about the secret and James chooses $\bs_J$ independently from the information sent on $\setE_1$ and from the randomness. Equation~\eqref{eq:secret_uni} holds because from the user's perspective the secret is drawn uniformly at random from $\F_q^{z_{wo}\alpha}$.

This equality implies that the user cannot distinguish whether $\setE_2$ or $\setE_3$ are sending information belonging to the true secret. The user must therefore decode the secret using only $X_{\setE_1}(\bs)$, otherwise the user would be making an error with probability $1/2(1-1/q^{z_{wo}\alpha})$. However, due to the privacy constraint any set of less than $z_{ro}+z_{rw}$ shares contain no information about the secret. Therefore, the capacity of the system is $0$. 
A similar argument follows if $k< 2z_{rw} + 2 z_{wo} + z_{ro}$. The adversary sends no information on $\setB \subset \setE_2$ of size $|\setE_2|-|\setE_3|$ links and the rest follows.  

\noindent{\em Case 2:} This case is straightforward. 
Assume that the dealer is storing a secret of maximal size $H(S) = (k- 2z_{rw} - z_{ro} - z_{wo})\alpha> z_{wo}\alpha$. James can choose a fake secret $\bs_J$ of size at most $z_{wo} = |\setE_2|-z_{rw}$. The user observes $X_{\setE_1}(\bs)$, $X_{\setE_2}(\bs_J)$ and $X_{\setE_3}(\bs)$. Due to privacy constraints, all true messages are equally likely from James' perspective. Thus, Jame's best strategy is to pick $\bs_J$ at random. Since James observes the information sent on the links of $\setE_1$, the information $X_{\setE_2}(\bs_J)$ will be consistent with $X_{\setE_1}(\bs)$. However, with high probability $X_{\setE_2}(\bs_J)$ is not consistent with $X_{\setE_3}(\bs)$ for all $\bs_J\neq \bs$. Therefore, the best the user can do is to detect that the information sent on $X_{\setE_2}(\bs_J)$ is independent from the true secret $\bs$ and use the remaining information to decode.

Hence, the amount of information the user can use to decode $\bs$ is upper bounded by the amount of information sent on $\setE_1$ and $\setE_3$. Due to privacy constraints, any collection of $z_{ro}+z_{rw}$ parties obtain no information about the secret. Thus, the capacity of the system is upper bounded by
\begin{equation*}
C \leq |\setE_1|+|\setE_3| - z_{ro} - z_{rw} = (k- 2z_{rw} - z_{ro} - z_{wo})\alpha.
\end{equation*}

Note that in this case the user knows that $X_{\setE_2}(\bs_J)$ is the set of corrupted parties with high probability. An error occurs if there exists a subset $\setA\subset \setE_3$ of size $H(S) -z_{rw}- z_{wo}$ links such that $X_{\setE_2}(\bs_J)\neq X_{\setA}(\bs)$ are both consistent with $X_{\setE_3\setminus \setA}(\bs)$. In this case, the user cannot decide whether $X_{\setE_2}(\bs_J)$ or $X_{\setA}(\bs)$ are information that belong to the true secret because all secrets are also equally distributed from his perspective. The best strategy of the user here is to decode only from $\setE_1$ and $\setE_3\setminus\setA$. Hence, the user cannot decode the secret if $(k- 3z_{rw} -2z_{wo} - z_{ro})\alpha<H(S)$ because out of the $|\setE_1|+|\setE_3|-|\setA| = k- 2z_{rw} -2z_{wo}$ links any $z_{ro}+z_{rw}$ links cannot have any information about the secret. This case happens with probability bounded by $q^{-H(S)+z_{wo}+z_{rw}}$. If the user wants to always decode the right message, then the capacity becomes the same as the capacity of error correction codes with privacy constraints, \ie $H(S)\leq (k- 2(z_{rw}+z_{wo}) - (z_{ro}+r_{rw}))\alpha$.

\paragraph{Minimum Download Cost}
To obtain an upper bound on the the minimum communication cost between any $d\leq n $ parties and the user we assume that each party can communicate at most $\beta\leq \alpha$ units of information to the user. We consider the case where $k>2z_{rw} + 2 z_{wo} + z_{ro}$. The user downloads $d\beta$ units of information from the parties. In a similar argument to the one used in the converse, the user cannot use the information sent from the $z_{wo}+z_{rw}$ parties because they send information independent from the secret $\bs$. Again, from the privacy constraint the information stored on (therefore sent from) any collection of $z_{ro}+z_{rw}$ parties does not contain any information about the secret $\bs$. Thus, the user can only use the remaining $(d-2z_{rw} - z_{ro} - z_{wo})\beta$ units of information to decode the secret. The useful information must be at least equal to size of the secret therefore
\begin{align}
(d-2z_{rw} - z_{ro} - z_{wo})\beta & \leq (k-2z_{rw} - z_{ro} - z_{wo})\alpha, \nonumber\\
\CC(d) = d\beta &\leq d \dfrac{(k-2z_{rw} - z_{ro} - z_{wo})\alpha}{d-2z_{rw} - z_{ro} - z_{wo}}\label{eq:ccproof}
\end{align}

We defer the proof using information theoretic inequalities, following the same steps of Section~\ref{sec:bound}, to the appendix.

\subsection{Omniscient Adversary}
\paragraph{Capacity} We quantify the amount of information that the dealer can send to a user contacting any $k$ parties. We do so by finding a cut in the network between the dealer and such a user. By the min-cut max-flow argument the capacity of the system is upper bounded by the value of this cut. We partition the contacted parties into three disjoint sets. Let $\setW$, $\setH_1$ and $\setH_2$ be three disjoint subsets of $\{1,\dots,n\}$ such that $|\setW|=z_{rw}+z_{wo}$, $|\setH_1|= z_{rw} + z_{wo}$ and $|\setH_2|=k- 2(z_{rw} - z_{wo)}$. We denote by $P_{\setW}$, $P_{\setH_1}$, $P_{\setH_2}$ the set of parties indexed by $\setR$, ${\setW}$ and ${\setH}$, respectively. A cut between the dealer and the user is $\setV_2 = \{U,P_{\setW}^\text{out}, P_{\setH_1}^\text{out}, P_{\setH_2}^\text{out}\}$ and $\setV_2 = \setV \setminus \setV_1$. The value of this cut is $k\alpha$.

Let $X_{\setW}(\bs), \ X_{\setH_1}(\bs), \ X_{\setH_2}(\bs)$ be the symbols sent to the user from the set of parties indexing $X$ when a secret $\mathbf{s}$ is being transmitted from the dealer. The adversary decides on a given secret $\bs_{J}$ of his choice independently from the true secret being sent by the dealer and sends $X_{\setW}(\bs_J)$ on the set of links he controls. Since the adversary is omniscient, he can always choose a secret $\bs_J$ such that both $X_{\setW}(\bs_J)$ and $X_{\setH_1}(\bs)$ are consistent with $X_{\setH_2}(\bs)$. The user cannot decide wether $\bs$ or $\bs_J$ is the true message and must decode the message using only $X_{\setH_2}(\bs)$. We formalize this intuition as follows. We assume that the secret $\bs$ is uniformly distributed over $\F_q^{H(S)}$. Let $\br$ and $\br_J$ be the values of the random numbers chosen by the dealer and the adversary respectively. We can write the following.
\begin{align}
\Pr \left( X_{\setW}(\bs_J) = x_1, X_{\setH_1}(\bs) = x_2  | X_{\setH_2}(\bs) = x_3 \right) & = \Pr \left (X_{\setH_1}(\bs) = x_2\right) \Pr \left (X_{\setW}(\bs_J) = x_1| X_{\setH_1}(\bs) = x_2, X_{\setH_2}(\bs) = x_3\right) \nonumber \\
& = \Pr \left(S = \bs, R=\br   \right) \Pr \left(S = \bs_J, R=\br_J | X_{\setH_1}(\bs) = x_2, X_{\setH_2}(\bs) = x_3  \right) \nonumber \\
& = \Pr \left(S = \bs \right)  \Pr \left(R = \br \right) \Pr \left(S = \bs_J\right) \Pr \left(R = \br_J|X_{\setH_1}(\bs) = x_2, X_{\setH_2}(\bs) = x_3   \right) \nonumber \\
& = \dfrac{1}{q^{H(S)}}  \Pr \left(R = \br \right) \dfrac{1}{q^{H(S)}}  \Pr \left(R = \br_J|X_{\setH_1}(\bs) = x_2, X_{\setH_2}(\bs) = x_3  \right) \label{eq:osecret_uni}\\
& =  \Pr \left(S = \bs_J \right)  \Pr \left(R = \br \right) \Pr \left(S = \bs \right) \Pr \left(R = \br_J| X_{\setH_1}(\bs) = x_2, X_{\setH_2}(\bs) = x_3  \right) \nonumber \\
& = \Pr \left( X_{\setW}(\bs) = x_1, X_{\setH_1}(\bs_J) = x_2  | X_{\setH_2}(\bs) = x_3 \right). \nonumber
\end{align}

Equation~\eqref{eq:osecret_uni} holds because from the user's perspective the secret is drawn uniformly at random from $\F_q^{H(S)}$. This equality implies that the user cannot distinguish whether $\setW$ or $\setH_1$ are sending information belonging to the true secret. The user must therefore decode the secret using only $X_{\setH_2}(\bs)$, otherwise the user would be making an error with probability $1/2$. However, due to the privacy constraint any set of less than $z_{ro}+z_{rw}$ shares contain no information about the secret. Therefore, the capacity of the system is less than or equal to $\left(|\setH_2|- z_{ro}+z_{rw}\right)\alpha = (k- 3z_{rw} - 2z_{wo} - z_{ro})\alpha$.

\paragraph{Minimum download} We follow the same argument above. Consider a user contacting $d$ parties and consider the trivial cut $\setV_2 = \{U\}$, $\setV_1 = \setV\setminus \setV_2$. Each party sends $\beta$ units of information to the dealer, which we want to minimize. The value of the cut is $d\beta$. Following the same reasoning as above, we know that the user can use at most $(d- 3z_{rw} - 2z_{wo} - z_{ro})\beta$ units of information. In order to decode the secret the total amount of useful downloaded information must be greater than or equal to $H(S) = (k- 3z_{rw} - 2z_{wo} - z_{ro})\alpha$. Therefore, we obtain
\begin{align*}
(d- 3z_{rw} - 2z_{wo} - z_{ro})\beta & \geq (k- 3z_{rw} - 2z_{wo} - z_{ro})\alpha, \\
\CC(d) = d\beta & \geq d\dfrac{ (k- 3z_{rw} - 2z_{wo} - z_{ro})\alpha}{d- 3z_{rw} - 2z_{wo} - z_{ro}}.
\end{align*}

\section{Achievability}

\subsection{Limited Knowledge Adversary}
\subsubsection{Capacity achieving codes}
To achieve capacity and minimum communication cost we use Staircase codes \cite{BRIT18} with a pairwise hash \cite{yao2014network,PRR11} added to the data given to the parties. Staircase codes is a family of communication efficient secret sharing that achieves capacity and minimum communication cost for all values of $k\leq d \leq n$ when $z_{w} = z_{wo}+z_{rw} = 0$. To construct an $(n,k,\bz)$ R-CE-SS, we need an $(n,k' = k-z_w, z'=z_r)$ Staircase code that achieves the minimum communication cost for $d'\in \{k',\dots,n-z_w\}$. The idea is for the user to contact $d=d'+z_w$ parties, use the pairwise hash to detect which $z_w$ parties are sending corrupted information and decode from the remaining parties. Note that from Staircase codes we get $$H(S) = (k'-z_r)\alpha = (k-z_w-z_r)\alpha = (k-2z_{rw}- z_{wo}- z_{ro})\alpha$$ and when contacting $d'$ parties each party sends $$\dfrac{CC(d')}{d'} = \dfrac{(k'-z_r)\alpha}{d'-z_r} = \dfrac{(k-2z_{rw}- z_{wo}- z_{ro})\alpha}{d-2z_{rw}- z_{wo}- z_{ro}}$$ units of information. Therefore, the capacity, privacy constraints and minimum communication cost and read costs are achieved. We need to prove that the hash can be used to catch the corrupted parties with high probability. 

The secret $\bs$ is a symbol drawn from a finite field $\F_q^\alpha$ for a power of a prime $q$ and an integer parameter $\alpha\triangleq \text{LCM}(n-z_w-z_r, \dots, k-z_w-z_r+1)$ of Staircase codes. Staircase code requires dividing each share $\bw_i$ into $\alpha$ symbols, $\bw_{i1},\dots, \bw_{i\alpha}$. We group the symbols of $\bw_i$ into $n-k+1$ vectors each of size $\gamma_\ell \triangleq \alpha_{k+\ell-1} - \alpha_{k+\ell}$ for $\ell=n-k+1,\dots,0$ (take $\alpha_{n+1} = 0$), such that a user contacting $d$ parties downloads the first group of symbols $\bw_{i\ell}$, $\ell = n-k+1,\dots, d-k+1$, of size $\alpha_d$ from each party. We view each symbol $\bw_{i\ell}$ as a vector over some finite field $\F_{q_1}^{\gamma_\ell}$. We construct the following hashes $h_{i\ell} = (\langle \bw_{i\ell}, \bw_{j\ell})$ for all $i \neq j \in \{1,\dots,n\}$. Each party $i$ stores $\alpha$ symbols $\bw_{i\ell}$ and $\alpha$ hash symbols $\bh_{i\ell}$ for $\ell=n-k+1,\dots,0$. Note that $\bw_{i\ell} \in\F_{q_1^v}$, whereas $\bh_\ell \in \F_{q_1}$ which can be made arbitrarily small. Therefore we assume that the size of the hash is negligible compared to the size of the secret.

\begin{table}[t!]
\renewcommand{\arraystretch}{1.2}
\centering
\begin{tabular}{l|c|c|c|c|c|c|c|c|c}
  \multicolumn{1}{c}{~}   &      \multicolumn{3}{c}{\small \em Read only}           & \multicolumn{3}{c}{\small \em Write}   & \multicolumn{3}{c}{\small \em Honest}   \\
  \multicolumn{1}{c}{~}   &      \multicolumn{3}{l}{\raisebox{0ex}{\smash{$\overbrace{\makebox[9em]{}}_{}$}}} & \multicolumn{3}{l}{\raisebox{0ex}{\smash{$\overbrace{\makebox[15em]{}}_{}$}}} & \multicolumn{3}{l}{\raisebox{0ex}{\smash{$\overbrace{\makebox[18em]{}}_{}$}}} \\
~ & $\hat{\bh}_{1}$ & $\cdots $& $\hat{\bh}_{z_{ro}}$ & $\hat{\bh}_{z_{ro}+1}$ & $\cdots$ & $\hat{\bh}_{z_{ro}+z_{rw}+z_{wo}}$&  $\hat{\bh}_{z_{ro}+z_{rw}+z_{wo} +1}$&$\cdots$ & $\hat{\bh}_{d}$\\ \hline
${\bh_{1}}$ & $\checkmark$ & $\cdots$ & \checkmark &  $\checkmark$ & $\cdots$ & $\checkmark$ & $\checkmark$ &  $\cdots$ &$\checkmark$ \\ \hline
$\vdots$ & $\vdots$ &$\vdots$  & $\vdots$ &  $\vdots$ & $\vdots$ &$\vdots$  & $\vdots$ &  $\vdots$ & $\vdots$  \\  \hline
${\bh_{z_{ro}}}$ & $\checkmark$ & $\cdots$ & \checkmark &  $\checkmark$ & $\cdots$ & $\checkmark$ & $\checkmark$ & $\cdots$ &$\checkmark$ \\ \hline

${\bh_{z_{ro}+1}}$& $\checkmark$ & $\cdots$ & \checkmark &  $\checkmark$ & $\cdots$ & $\checkmark$ & \cellcolor{green!25}$\times$ &\cellcolor{green!25} $\cdots$ & \cellcolor{green!25} $\times$ \\ \hline
$\vdots$ & $\vdots$ &$\vdots$  & $\vdots$ &  $\vdots$ & $\vdots$ &$\vdots$  & \cellcolor{green!25}$\vdots$ &  \cellcolor{green!25}$\vdots$ &\cellcolor{green!25} $\vdots$  \\  \hline
${\bh_{z_{ro}+z_{rw}+z_{wo}}}$& $\checkmark$ & $\cdots$ & \checkmark &  $\checkmark$ & $\cdots$ & $\checkmark$ &\cellcolor{green!25} $\times$ & \cellcolor{green!25}$\cdots$ & \cellcolor{green!25}$\times$ \\ \hline

${\bh_{z_{ro}+z_{rw}+z_{wo} +1} }$& $\checkmark$ & $\cdots$ & \checkmark & \cellcolor{red!25} $\times$ &\cellcolor{red!25} $\cdots$ & \cellcolor{red!25}$\times$ & $\checkmark$ & $\cdots$ &$\checkmark$ \\ \hline
$\vdots$ & $\vdots$ &$\vdots$  & $\vdots$ &  \cellcolor{red!25}$\vdots$ & \cellcolor{red!25}$\vdots$ &\cellcolor{red!25} $\vdots$  & $\vdots$ &  $\vdots$ & $\vdots$  \\  \hline
${\bh_{d} }$& $\checkmark$ & $\cdots$ & \checkmark & \cellcolor{red!25} $\times$ & \cellcolor{red!25}$\cdots$ & \cellcolor{red!25} $\times$ & $\checkmark$ & $\cdots$ &$\checkmark$ \\  \hline
\end{tabular}
\caption{The hash comparison table assuming that the first $z_{ro}+z_{wo}+z_{rw}$ parties are controlled by the adversary.}
\label{tab:hashgeneral}
\end{table}

Assume for now that James can neither observe nor corrupt the hash. We will show in the next Section how the hash can be made secure and private from James. A user contacting $d$ parties, reads and downloads the first $\alpha_d\triangleq\dfrac{(k-2z_{rw}- z_{wo}- z_{ro})\alpha}{d-2z_{rw}- z_{wo}- z_{ro}}$ symbols from each share and the corresponding hashes. Assume without loss of generality that the first $z_{wo}+z_{rw}$ parties are controlled by the adversary. The user concatenates the downloaded hashes of each share into one vector $\bh_i$ and computes $\hat{\bh}_i$ using the shares he downloaded to constructs the hash comparison table given in Table~\ref{tab:hashgeneral}. A $\checkmark$ means that the computed hash matches the downloaded hash and a $\times$ means the opposite.

As long as the number of columns in the green part of the table is greater than the number of columns in the red part the user can successfully detect which parties are sending corrupted data. The number of columns in the green part is $d-z_{rw}-z_{wo}- z_{ro}$ and the number of columns in the red part is $z_{rw}+ z_{wo}$. Therefore, the adversary has to introduce errors orthogonal to at least $f=d-2z_{rw}-2z_{wo}- z_{ro}>1$ columns. Thus the probability of error is bounded by
\begin{align*}
\Pr(\text{error}) &= \Pr(\text{flipping  $f$ ``$\times$'' to ``$\checkmark$''})\\
& = \Pr(\mathbf{e} \bot \bw_{i_1}) \Pr(\mathbf{e} \bot \bw_{i_2}| \mathbf{e} \bot \bw_{i_1}) \cdots \Pr(\mathbf{e} \bot \bw_{i_f}| \mathbf{e} \bot \bw_{i_1}, \dots,  \mathbf{e} \bot \bw_{i_{f-1}})\\
& \leq \Pr(\mathbf{e} \bot \bw_{i_1})\\
&\stackrel{(a)}{=} \left(\dfrac{q_1^{\gamma_\ell -1}}{q_1^{\gamma_\ell}} \right)^{n-d+1}\\
& = \dfrac{1}{q_1^{n-d+1}}.
\end{align*}

The equality (a) follows because by construction each $w_i$ consists of the concatenation of $n-d+1$ independent vectors of size $\gamma_\ell$ each. Therefore the adversary wants to find $n-d+1$ error vectors, each being orthogonal to the corresponding part of $\bw_i$. The probability of finding such a vector is the probability of finding a vector that lies in the space of dimension $q^{\gamma_\ell -1}$ orthogonal to the corresponding part of $\bw_i$.

\subsubsection{Securely Storing the Hash}
We show how to store the has securely and privately from the adversary and analyze the overhead required to store the hash. Recall that each party stores $n-k+1$ hash vectors $\bh_{i\ell}$, $\ell = n-k+1,\dots, 0$. We slightly modify the creation of the hash. For each value of $\ell$ we create $\bh_{\ell} = (\langle \bw_{i\ell}, \bw_{j\ell}\rangle)$ for all $n\neq j \in \{1,\dots,n\}$. Note that this $\bh_{\ell}$ is indeed the concatenation of the hashes created in the previous section. We encode each $\bh_{\ell}$ using an $(n,k_h = z_{ro}+z_{rw}+1,z_h=z_{ro}+z_{rw})$ secret sharing code and distributed the resulting shares to the parties.

\paragraph{Privacy and security of the hash}
As a result of the use of secret sharing the values of the hash are private from an adversary observing any $z_r=z_h$ shares. We want to show that a user contacting $d$ parties can obtain the correct values of the hash vectors. The user downloads any $k_h$ shares of the hash secret sharing code. Out of these $k_h$ shares at most $z_w = z_{wo}+ z_{rw}$ values are corrupted by James. The minimum distance of the hash secret sharing is $d_{\min} = n-k_h+1 = n-z_{ro}-z_{rw}+1 > z_{rw}+z_{wo} + 2$, because $n\geq k>2z_{rw}+z_{ro}+z_{wo}$. Therefore, as a property of secret sharing, which are maximum distance separable codes, the user can detect any $d_{\min} - 1 > z_{rw}+z_{wo} + 1$ errors. To that end, the user checks the $\binom{d}{k_h}$ possible values of the hash shares and decodes from a set that has no errors. Note that since $d\geq k > 2z_{rw}+z_{ro}+z_{wo} = k_h + z_{rw}+z_{wo} - 1$ there is always a set of $k_h$ uncorrupted hash shares that the user can decode from.

\paragraph{Rate analysis}
We analyze the overhead of storing the hash on the parties. Each vector $\bh_{\ell}$ is of length $\gamma_\ell n(n-1)/2$ and so is the length of each hash secret share. The total length of the hash secret shares stored on one parties is $\sum_{\ell} \gamma_\ell n(n-1)/2 = \alpha n(n-1)/2$ over $\F_{q_1}$. Therefore, the overhead of the hash vectors is equal to
\begin{align*}
\log\left(\dfrac{q_1^{\alpha n (n-1)/2}}{q_1^{v\alpha}}\right) = \dfrac{n(n-1)}{2v}.
\end{align*}

The overhead of the hash can be made arbitrarily small by increasing $v$.

\paragraph{Reducing the hash overhead}
While computing the hash values we assumed that we want to cross check the information sent from party $i$ with all the other $n-1$ parties so that the user can find a set of $k-z_w$ parties that have sent consistent information to the user. This assumption lead to the overhead of hash being $n(n-1)/2$. 

First observe that we can model this problem using a graph. The parties are represented using $n$ vertices. An edge is drawn between a pair of vertices $i$ and $j$ if there exists a hash value $\bh_{i,j}$ comparing the consistency of the information sent from parties $i$ and $j$. In our initial setting we assumed that the graph is a complete graph. When downloading information from $d$ parties, the user looks at the subgraph induced by the vertices corresponding to the contacted parties. The user compares the computed hashes to the downloaded hashes and deletes all the edges where the computed hash is different from the downloaded hash. The goal of the user is to find a connected component of size $d-z_{w}$ representing the parties that sent uncorrupted information. We showed that this is possible with high probability when the graph is a complete graph.

To reduce the overhead of the hash, notice that the graph on $n$ vertices must only be connected. Therefore, a user contacting any $d$ parties can look at the induced subgraph on the $d$ vertices corresponding to the contacted parties and repeat the same process described above. It can be shown that a graph on $n$ vertices is connected almost surely if a vertex $i$ is connected to a vertex $j\neq i$, $j\in \{1,\dots,n\},$ with probability $p=(\log(n) + \log(\log(n)))/n$. Let $\mathcal{N}(i)$ be the set of vertices connected to vertex $i$, we have $\E|\mathcal{N}| = \log(n) + \log(\log(n))$. Thus, in our setting, each hash vector becomes $\bh_\ell = (\langle \bw_i, \bw_j \rangle)$ for all $i=1,\dots,n$, $j=\mathcal{N}(i)$. This reduces the hash overhead to $\dfrac{n(\log (n)+\log(\log(n))}{v}.$
 
\subsection{Omniscient Adversary}
Capacity achieving codes that achieve minimum communication and read costs for this case are based on Staircase codes \cite{BRIT18}. To construct an $(n,k,\bz)$ R-CE-SS code, we need an $(n,k'=k-2z_w, z=z_r)$ Staircase code. that achieves minimum read and communication costs for all $d'\in \Delta$ where $\Delta = \{k-2z_w,\dots, n-2z_w\}$. The idea is for the user to contact $d=d'+2z_w$ parties and correct the worst case errors. Note that from Staircase codes we get $$H(S) = (k'-z_r)\alpha = (k-2z_w-z_r)\alpha = (k-3z_{rw}-2 z_{wo}- z_{ro})\alpha$$ and when contacting $d'$ parties each party sends $$\dfrac{CC(d')}{d'} = \dfrac{(k'-z_r)\alpha}{d'-z_r} = \dfrac{(k-3z_{rw}- 2z_{wo}- z_{ro})\alpha}{d-3z_{rw}-2 z_{wo}- z_{ro}}$$ units of information. Therefore, the capacity, privacy constraints and minimum communication cost and read costs are achieved. Since Staircase codes can be viewed as a collection of concatenated Reed-Solomon codes, the error correction capability of those codes follow immediately from the error correction capability of Reed-Solomon codes.

\section{Conclusion}
We studied communication efficient secret sharing schemes that tolerate the presence of malicious parties trying to actively corrupt the data stored in the distributed system. We assume the knowledge of the adversary is restricted to the information given to the compromised parties. We show that leveraging the limitation of the knowledge of the adversary allows an increase in the size of the shared secret. We use flow graph representation of the R-CE-SS setting to characterize the capacity, \ie maximum size of the secret that can be shared, when the adversary has limited and full knowledge of the shared information. We also characterize the minimum amount of information, as a function of the number of stragglers, needed to to be read and communicated to a legitimate user to decode the secret. We construct codes that achieve capacity in both settings. In addition, the constructed codes achieve minimum read and communication costs for any number of stragglers, up to a given threshold. Our codes are based on Staircase codes previously introduced for communication efficient secret sharing and the use of a pairwise hashing scheme used in distributed data storage and network coding settings to detect the presence of a limited knowledge adversary.

\noindent{\em Open problem:} Our main motivation for studying secret sharing stems from its application to several other settings. One can prove that the capacity of distributed storage system, private information retrieval and minimum communication cost of distributed computing are the same. However, it is not clear how to extended R-CE-SS codes to work in the other settings. For instance, in private information retrieval, the hash used by the user must be a function of the stored data (to be retrieved) as well as the encoding scheme used to retrieve the data. A similar argument holds for distributed computing.

\bibliographystyle{IEEEtran}
\bibliography{Thesis}

 \appendix
We denote the data sent from party $i$ to the user by $\mathbf{y}_i$ and the corresponding random variable by $Y_i$. Since the user must be able to decode the secret from any collection of $Y_i$ of size $d$, we have $H(S|Y_1^d) =  0$. Following the same steps of Section~\ref{sec:bound} we write
\begin{align}
H(S) &=  I(S; Y_1^d) \label{eq:dmutual}\\
 & = I(S; Y_1^{z_{w}}) + I(S; Y_{z_{w}+1}^{d}| Y_1^{z_{w}}) \label{eq:dchain}\\
 & = I(S; Y_{z_{w}+1}^{d}| Y_1^{z_{w}}) \label{eq:dpri}\\
 & = H(Y_{z_{w}+1}^{d}| Y_1^{z_{w}}) - H(Y_{z_{w}+1}^{d} | S,Y_1^{z_{w}}) \nonumber\\
 & = H(Y_{z_{w}+z_{r}+1}^{d}| Y_1^{z_{w}}) \nonumber \\
 & ~~ +  H(Y_{z_{w}+1}^{z_{w}+z_{r}}| Y_1^{z_{w}}, Y_{z_{w}+z_{r}+1}^{d}) - H(Y_{z_{w}+1}^{d} | S,Y_1^{z_{w}}) \label{eq:dchain2} \\
 & \leq H(Y_{z_{w}+z_{r}+1}^{d}) \label{eq:ddpi} \\
 & \leq (d-2z_{rw}-z_{wo} - z_{ro})H(Y_i) \label{eq:dind}\\
 & = (d-2z_{rw}-z_{wo} - z_{ro})\beta. \label{eq:beta}
\end{align}

Equation~\eqref{eq:dchain} follows from the chain rule of mutual information. Equation~\eqref{eq:dpri} follows from the privacy constraint given in~\eqref{eq:privacy}. In~\eqref{eq:dpri} we removed the first $z_{ro}$ shares which does not incur loss of generality. Equation~\eqref{eq:dchain2} follows from the chain rule of entropy. Equation~\eqref{eq:ddpi} follows from the data processing inequality as we shall show next. Equation~\eqref{eq:dind} follows from the chain rule of entropy and~\eqref{eq:beta} follows because $H(Y_i) = \beta$. 

To show that~\eqref{eq:ddpi} holds we use the non-negativity of the entropy and write the following.
\begin{align}
H(Y_{z_{w}+1}^{d} | S,Y_1^{z_{w}}) & = H(Y_{z_{w}+1}^{z_{w}+z_{r}}| S,Y_1^{z_{w}}) +   H(Y_{z_{w}+z_{r}}^{d}| S,Y_1^{z_{w}}, Y_{z_{w}+1}^{z_{w}+z_{r}}) \nonumber \\
& \geq H(Y_{z_{w}+1}^{z_{w}+ z_{r}}| S,Y_1^{z_{w}}).\label{eq:dgreater}
\end{align}

Let $$Q\triangleq H(Y_{z_{w}+1}^{z_{w}+z_{r}}| Y_1^{z_{w}},Y_{z_{w}+z_{r}+1}^{d}) - H(Y_{z_{w}+1}^{d} | S,Y_1^{z_{w}}),$$ we use~\eqref{eq:dgreater} to bound $Q$ as follows.
\begin{align}
Q &\leq H(Y_{z_{w}+1}^{z_{w}+z_{r}}| Y_1^{z_{w}},Y_{z_{w}+ z_{r}+1}^{d}) - H(Y_{z_{w}+ 1}^{z_{w}+z_{r}}| S,Y_1^{z_{w}}) \nonumber \\
& =I(Y_{z_{w}+1}^{z_{w}+z_{r}}; S,Y_1^{z_{w}}) - I(Y_{z_{w}+1}^{z_{w}+z_{r}}; Y_1^{z_{w}},Y_{z_{w}+z_{r}+1}^{d})\label{eq:ddefi}\\
& \geq 0.\label{eq:ddpi1}
\end{align}

Equation~\eqref{eq:ddefi} follows from the definition of mutual information $I(A;B) = H(A) - H(A|B)$ and~\eqref{eq:ddpi1} holds because $S\to W_{z_{w}+z_{r}+1}^{d}\to Y_{z_{w}+z_{r}+1}^{d}$ forms a Markov chain and we can use the data processing inequality.
\end{document}

%% file: rawad_macros.tex
\usepackage[table]{xcolor}

\usepackage{%
	graphicx,%
	array,%
	caption,%
	cite,%
	diagbox,%
	epsfig,%
	latexsym,%
	multirow,%
	pgfplots,%
	pstricks,
	sidecap,%
	subcaption,
	varwidth,%
	wrapfig,%
	arydshln,%
	slashbox,%
	nccmath,%
	float,%
	algorithmic,%
	epstopdf,%
	bm,%
	bbm,%
	nameref,%
	bchart,%
	nicefrac,%
	comment,
	tikz,%
	tikzpeople,%
	mathtools,%
}

\usepackage[T1]{fontenc}

\usepackage[inline]{enumitem}
\usepackage[linesnumbered,ruled,vlined]{algorithm2e}

\usepackage{amssymb,amsmath,amsthm,amsfonts,accents}

\definecolor{DarkGreen}{rgb}{0.1,0.5,0.1}
\definecolor{DarkRed}{rgb}{0.5,0.1,0.1}
\definecolor{DarkBlue}{rgb}{0.1,0.1,0.5}

\usepackage[pdftex]{hyperref}
\hypersetup{
    unicode=false,          
    pdftoolbar=true,        
    pdfmenubar=true,        
    pdffitwindow=false,      
    pdfnewwindow=true,      
    colorlinks=true,       
    linkcolor=DarkBlue,          
    citecolor=DarkGreen,        
    filecolor=DarkGreen,      
    urlcolor=DarkBlue,          
    breaklinks=true,		
    pdftitle={},
    pdfauthor={},    
}

\usetikzlibrary{%
	arrows,%
	automata,%
	calc,%
	patterns,%
	plotmarks,%
	positioning,%
	shapes.geometric,%
	shapes,%
	snakes,%
	decorations.pathmorphing,%
	decorations.shapes,%
	graphs,%
	chains,%
}

\graphicspath{{Figures/}}

\pgfplotsset{
    legend image with text/.style={
        legend image code/.code={%
            \node[anchor=center] at (0.3cm,0cm) {#1};
        }
    },
}

\newcolumntype{C}[1]{>{\centering\let\newline\\\arraybackslash\hspace{0pt}}m{#1}}
\newcolumntype{L}[1]{>{\raggedright\let\newline\\\arraybackslash\hspace{0pt}}m{#1}}

\newcommand{\F}{{\mathbb F}}

\newcommand{\bs}{\mathbf{s}}
\newcommand{\bz}{\mathbf{z}}
\newcommand{\bw}{\mathbf{w}}
\newcommand{\bh}{\mathbf{h}}
\newcommand{\br}{\mathbf{r}}

\newcommand{\be}{\begin{equation}}
\newcommand{\ee}{\end{equation}}
\newcommand{\bear}{\begin{eqnarray}}
\newcommand{\eear}{\end{eqnarray}}
\newcommand{\bears}{\begin{eqnarray*}}
\newcommand{\eears}{\end{eqnarray*}}
\newcommand{\ben}{\begin{enumerate}}
\newcommand{\een}{\end{enumerate}}

\newcommand{\ie}{{i.e., }}

\definecolor{ogreen}{rgb}{0,0.5,0}
\definecolor{magenta}{rgb}{1.0, 0.11, 0.81}
\definecolor{mulberry}{rgb}{0.77, 0.29, 0.55}
\definecolor{xgray}{rgb}{0.9, 0.9, 0.9}
\definecolor{bittersweet}{rgb}{1.0, 0.44, 0.37}
\definecolor{darkterracotta}{rgb}{0.8, 0.31, 0.36}
\definecolor{darkturquoise}{rgb}{0.0, 0.81, 0.82}
\definecolor{chartreuse}{rgb}{0.87, 1.0, 0.0}
\definecolor{darkpastelgreen}{rgb}{0.01, 0.75, 0.24}
\definecolor{sandstorm}{rgb}{0.93, 0.84, 0.25}
\definecolor{rossocorsa}{rgb}{0.83, 0.0, 0.0}
\definecolor{bleudefrance}{rgb}{0.19, 0.55, 0.91}

\def \blue{\color{blue}}

\definecolor{byzantium}{rgb}{0, 0, 0.5}

\def \be{\begin{equation}}
\def \ee{\end{equation}}

\def \CC{{\fontfamily{qcr}\selectfont\text{CC}}}
\def \RC{{\fontfamily{qcr}\selectfont\text{RC}}}

\def \setA {\mathcal{A}}
\def \setB {\mathcal{B}}
\def \setE {\mathcal{E}}
\def \setH {\mathcal{H}}
\def \setR {\mathcal{R}}
\def \setI {\mathcal{I}}
\def \setV {\mathcal{V}}
\def \setW {\mathcal{W}}
\def \setZ {\mathcal{Z}}

\def \bes{\begin{equation*}}
\def \ees{\end{equation*}}
\def \bas{\begin{align*}}
\def \eas{\end{align*}}
\def \bbm{\begin{bmatrix}}
\def \ebm{\end{bmatrix}}

\newcommand{\E}{\mathbb{E}}

\renewcommand{\epsilon}{\varepsilon}

\theoremstyle{plain}
\newtheorem{example}{Example}
\newtheorem{theorem}{Theorem}
\newtheorem*{theorem*}{Theorem}

\theoremstyle{definition}

\theoremstyle{remark}

%% file: Comm-Eff_SS_with_LK.bbl
\begin{thebibliography}{10}
\providecommand{\url}[1]{#1}
\csname url@samestyle\endcsname
\providecommand{\newblock}{\relax}
\providecommand{\bibinfo}[2]{#2}
\providecommand{\BIBentrySTDinterwordspacing}{\spaceskip=0pt\relax}
\providecommand{\BIBentryALTinterwordstretchfactor}{4}
\providecommand{\BIBentryALTinterwordspacing}{\spaceskip=\fontdimen2\font plus
\BIBentryALTinterwordstretchfactor\fontdimen3\font minus
  \fontdimen4\font\relax}
\providecommand{\BIBforeignlanguage}[2]{{%
\expandafter\ifx\csname l@#1\endcsname\relax
\typeout{** WARNING: IEEEtran.bst: No hyphenation pattern has been}%
\typeout{** loaded for the language `#1'. Using the pattern for}%
\typeout{** the default language instead.}%
\else
\language=\csname l@#1\endcsname
\fi
#2}}
\providecommand{\BIBdecl}{\relax}
\BIBdecl

\bibitem{S79}
A.~Shamir, ``How to share a secret,'' \emph{Communications of the ACM},
  vol.~22, no.~11, pp. 612--613, 1979.

\bibitem{McESa81}
R.~J. McEliece and D.~V. Sarwate, ``On sharing secrets and reed-solomon
  codes,'' \emph{Communications of the ACM}, vol.~24, no.~9, pp. 583--584,
  1981.

\bibitem{PRR11}
S.~Pawar, S.~El~Rouayheb, and K.~Ramchandran, ``Securing dynamic distributed
  storage systems against eavesdropping and adversarial attacks,'' \emph{IEEE
  Transactions on Information Theory}, vol.~57, no.~10, pp. 6734--6753, 2011.

\bibitem{AF10}
M.~J. Atallah and K.~B. Frikken, ``Securely outsourcing linear algebra
  computations,'' in \emph{Proceedings of the 5th ACM Symposium on Information,
  Computer and Communications Security (ASIACCS)}, 2010, pp. 48--59.

\bibitem{BPR17}
R.~Bitar, P.~Parag, and S.~El~Rouayheb, ``Minimizing latency for secure
  distributed computing,'' in \emph{IEEE International Symposium on Information
  Theory (ISIT)}, 2017, pp. 2900--2904.

\bibitem{chor1998private}
B.~Chor, E.~Kushilevitz, O.~Goldreich, and M.~Sudan, ``Private information
  retrieval,'' \emph{Journal of the ACM (JACM)}, vol.~45, no.~6, pp. 965--981,
  1998.

\bibitem{SMPCbook}
R.~Cramer, I.~B. Damgrd, and J.~B. Nielsen, \emph{Secure Multiparty Computation
  and Secret Sharing}, 1st~ed.\hskip 1em plus 0.5em minus 0.4em\relax New York,
  NY, USA: Cambridge University Press, 2015.

\bibitem{DB13}
J.~Dean and L.~A. Barroso, ``The tail at scale,'' \emph{Communications of the
  ACM}, vol.~56, no.~2, pp. 74--80, 2013.

\bibitem{WW08}
H.~Wang and D.~S. Wong, ``On secret reconstruction in secret sharing schemes,''
  \emph{IEEE Transactions on Information Theory}, vol.~54, no.~1, pp. 473--480,
  Jan 2008.

\bibitem{BRIT18}
R.~Bitar and S.~El~Rouayheb, ``Staircase codes for secret sharing with optimal
  communication and read overheads,'' \emph{IEEE Transactions on Information
  Theory}, vol.~64, no.~2, pp. 933--943, 2018.

\bibitem{HLKBtrans}
W.~Huang, M.~Langberg, J.~Kliewer, and J.~Bruck, ``Communication efficient
  secret sharing,'' \emph{IEEE Transactions on Information Theory}, vol.~62,
  no.~12, pp. 7195--7206, 2016.

\bibitem{JLKHKME08}
Jaggi, M.~Langberg, S.~Katti, T.~Ho, D.~Katabi, M.~Medard, and M.~Effros,
  ``Resilient network coding in the presence of {B}yzantine adversaries,''
  \emph{IEEE Transactions on Information Theory (special issue on
  information-theoretic security)}, pp. 2596--2603, 2008.

\bibitem{ZYSMH12}
Z.~Zhang, Y.~M. Chee, S.~Ling, M.~Liu, and H.~Wang, ``Threshold changeable
  secret sharing schemes revisited,'' \emph{Theoretical Computer Science}, vol.
  418, pp. 106--115, 2012.

\bibitem{yao2014network}
H.~Yao, D.~Silva, S.~Jaggi, and M.~Langberg, ``Network codes resilient to
  jamming and eavesdropping,'' \emph{IEEE/ACM Transactions on networking},
  vol.~22, no.~6, pp. 1978--1987, 2014.

\bibitem{bitar2015securing}
R.~Bitar and S.~El~Rouayheb, ``Securing data against limited-knowledge
  adversaries in distributed storage systems,'' in \emph{IEEE International
  Symposium on Information Theory (ISIT)}, 2015, pp. 2847--2851.

\bibitem{schaefer2017information}
R.~F. Schaefer, H.~Boche, A.~Khisti, and H.~V. Poor, \emph{Information
  Theoretic Security and Privacy of Information Systems}.\hskip 1em plus 0.5em
  minus 0.4em\relax Cambridge University Press, 2017.

\bibitem{DGWWR07}
A.~G. Dimakis, B.~Godfrey, Y.~Wu, M.~Wainwright, and K.~Ramchandran, ``Network
  coding for distributed storage systems,'' \emph{IEEE transactions on
  information theory}, vol.~56, no.~9, pp. 4539--4551, 2010.

\end{thebibliography}
